\documentclass[10pt, journal, twocolumn]{IEEEtran}
\IEEEoverridecommandlockouts

\usepackage{cite}
\usepackage{url}
\usepackage{multirow}
\usepackage{svg}
\usepackage{amsmath,amssymb,amsfonts}
\usepackage{booktabs}
\usepackage{siunitx} 
\usepackage{algorithm}
\usepackage{algpseudocode}
\usepackage{graphicx}
\graphicspath{{./figs/}}
\usepackage{xcolor}

\usepackage[caption=false,font=footnotesize]{subfig}
\def\BibTeX{{\rm B\kern-.05em{\sc i\kern-.025em b}\kern-.08em
    T\kern-.1667em\lower.7ex\hbox{E}\kern-.125emX}}
\usepackage{textcomp}
\usepackage{comment}
\usepackage{tikz}
\usepackage{pgfplots}
\pgfplotsset{compat=1.18}
\usetikzlibrary{patterns}
\usetikzlibrary{arrows,shapes,backgrounds,fit,calc}
\usetikzlibrary{automata, positioning, arrows.meta}
\usepackage[nolist]{acronym}
\usepackage[
    hidelinks,
    colorlinks=true,
    linkcolor=black,
    citecolor=black,
    urlcolor=black
]{hyperref}

\DeclareMathOperator*{\argmax}{\arg\!\max}

\floatname{algorithm}{Protocol}

\begin{document}

\title{
Blind and Asynchronous Initial Beam Acquisition and Clock-Assisted Re-Acquisition in Standalone Mobile MIMO-OFDM Systems
\\}

\author{Ali~Abdi, 
        Firooz~B.~Saghezchi,~\IEEEmembership{Senior~Member,~IEEE,}
        and~Haris~Gacanin,~\IEEEmembership{Fellow,~IEEE}
\thanks{A preliminary version of this work was submitted to the IEEE 105th Vehicular Technology Conference (VTC2027-Spring) \cite{abdi2026vtc}. This work was partially funded by the German Federal Ministry of Research, Technology and Space (BMFTR) through 6GEM+ (16KIS2409K), GEM-X (16KISS004K), and VICTOR6G (16KIS2547) projects.}
\thanks{The authors are with the Chair for Distributed Signal Processing, RWTH Aachen University, Aachen 52074, Germany (e-mail: abdi@dsp.rwth-aachen.de; saghezchi@dsp.rwth-aachen.de; harisg@dsp.rwth-aachen.de).}%
}

\maketitle



\acrodef{CSI}{Channel State Information}
\acrodef{ISAC}{Integrated Sensing and Communications}
\acrodef{GPS}{Global Positioning System}
\acrodef{mmWave}{Millimeter-Wave}
\acrodef{OFDM}{Orthogonal Frequency-Division Multiplexing}
\acrodef{5G}{Fifth-Generation}
\acrodef{PC}{Personal Computer}
\acrodef{SDR}{Software-Defined Radio}
\acrodef{RF}{Radio Frequency}
\acrodef{IF}{Intermediate Frequency}
\acrodef{BS}{Base Station}
\acrodef{UE}{User Equipment}
\acrodef{UPA}{Uniform Planar Array}
\acrodef{2D}{Two-Dimensional}
\acrodef{GRC}{GNU Radio Companion}
\acrodef{GUI}{Graphical User Interface}
\acrodef{AOD}{Angle of Departure}
\acrodef{AOA}{Angle of Arrival}
\acrodef{LoS}{Line of Sight}
\acrodef{NLoS}{Non-Line of Sight}
\acrodef{OOT}{Out-of-Tree}
\acrodef{DL}{Downlink}
\acrodef{UL}{Uplink}
\acrodef{TX}{Transmitter}
\acrodef{RX}{Receiver}
\acrodef{PHY}{Physical Layer}
\acrodef{HPBW}{Half-Power Beam-Width}
\acrodef{SNR}{Signal-to-Noise Ratio}
\acrodef{MU}{Multi-User}
\acrodef{SU}{Single-User}
\acrodef{MIMO}{Multiple-Input Multiple-Output}
\acrodef{SSB}{Synchronization Signal Block}
\acrodef{IA}{Initial Acquisition}
\acrodef{TDD}{Time Division Duplex}
\acrodef{USRP}{Universal Software Radio Peripheral}
\acrodef{NR}{New Radio}
\acrodef{RSS}{Received Signal Strength}
\acrodef{HW}{Hardware}
\acrodef{BIBA}{Blind Initial Beam Acquisition}
\acrodef{CARA}{Clock-Assisted Re-Acquisition}
\acrodef{FSM}{Finite State Machine}



\begin{abstract}
Building upon our preliminary study, this paper proposes a standalone beam-management framework for mobile MIMO-OFDM systems that integrates \ac{BIBA} with \ac{CARA}. \ac{BIBA} is a blind and asynchronous protocol that requires neither prior synchronization nor an out-of-band feedback channel, so the \ac{BS} and \ac{UE} can independently identify the beam pair maximizing received signal strength. To handle subsequent link drops, the newly introduced \ac{CARA} protocol exploits the clock synchronization established after \ac{BIBA} to reduce the re-acquisition time by approximately 55\% compared with \ac{BIBA}. The operational reliability of both protocols is bounded by a probabilistic feasibility analysis under random \ac{UE} mobility trajectories. Protocols are validated on a 28 GHz mmWave testbed under practical hardware constraints, including beam-switching and \ac{TDD} latencies, and measured \ac{SDR} software processing delays. Finally, we formulate a unified \ac{FSM} model for both \ac{BIBA} and \ac{CARA} to ensure robust integration into practical wireless systems.

\acresetall

\end{abstract}


\section{Introduction}
\label{sec:Introduction}
\IEEEPARstart{T}{o} enable standalone operation, mobile MIMO-OFDM systems with highly directive antennas (e.g., mmWave, Terahertz) must operate without relying on traditional \ac{IA} protocols that require pre-established synchronization or out-of-band feedback channels~\cite{aykin2020efficient, wei2019two}. Without these aids, the \ac{BS} and \ac{UE} are time-unsynchronized and lack shared spatial information (e.g., location or relative beam direction) or \ac{CSI}. Because beam misalignment yields \acp{SNR} too low for reliable time synchronization, \ac{IA} must be achieved blindly without prior spatial information, \ac{CSI}, or feedback channels. Only after the successful completion of \ac{IA} can sufficient link budget be established for reliable \ac{OFDM} frame synchronization~\cite{Schmidl1997}.

While recent two-phase \ac{DL}/\ac{UL} protocols aim to reduce beam acquisition overhead or eliminate feedback, they often rely on random beam probing, external sensing, or assumptions that neglect practical beam and \ac{TDD} switching latencies. These limitations motivate a fully blind, asynchronous beam acquisition protocol that requires no prior synchronization, spatial information, \ac{CSI}, or feedback while accounting for hardware switching latencies~\cite{danneberg2020usrp}.

In this paper, we bridge the gap between theoretical protocols and physical implementations by proposing a practical standalone beam-management framework that integrates blind \ac{IA} with clock-assisted re-acquisition. Building upon our preliminary work \cite{abdi2026vtc} which introduced the baseline \ac{BIBA} protocol, a fully blind and asynchronous initial beam acquisition protocol, this manuscript extends the architecture into a unified \ac{FSM}. Within this framework, \ac{BIBA} identifies the transmit-receive beam pair maximizing the \ac{RSS} without relying on prior synchronization, spatial information, \ac{CSI}, or out-of-band feedback channels. Once \ac{BIBA} establishes the initial link, the \ac{BS} and \ac{UE} synchronize their local clocks. Because this synchronization is retained even after a sudden physical link drop (e.g., due to user mobility), our framework subsequently exploits it by introducing the \ac{CARA} protocol. By leveraging the retained clock synchronization, \ac{CARA} schedules a rapid, deterministically aligned beam sweep that resolves the initial asynchronous timing ambiguity, reducing the overall re-acquisition time by approximately 55\%. Finally, we validate this unified framework on a standalone 28\,GHz \ac{SDR}-based testbed with phased-array antennas, demonstrating robustness under realistic \ac{TDD} switching latencies and hardware constraints.

Our contributions are summarized as follows:
\begin{itemize}
    \item We propose a comprehensive standalone beam-management framework. This framework incorporates our baseline \ac{BIBA} protocol \cite{abdi2026vtc} for fully blind, asynchronous initial beam acquisition, alongside the newly introduced \ac{CARA} protocol for rapid, clock-assisted beam re-acquisition.
    \item We derive closed-form expressions for the probability of feasible acquisition under random \ac{UE} mobility trajectories, establishing strict theoretical mobility boundaries (maximum speed and distance) to guarantee reliable operation.
    \item We validate the framework on a standalone 28\,GHz \ac{SDR} testbed, demonstrating practical viability under strict hardware constraints, specifically physical \ac{TDD} and beam switching latencies and measured \ac{SDR} software processing delays.
    \item We formulate a unified \ac{FSM} to govern the beam management process, ensuring seamless transitions between asynchronous initial beam acquisition, synchronized data transmission, and link recovery.
\end{itemize}

\section{Related Work}
\label{sec:Related_Works}
Table~\ref{tab:sota_comparison} summarizes the landscape of existing beam acquisition protocols and highlights the capability gaps addressed by our proposed framework. In the literature, several studies investigate reducing beam acquisition time through two-stage or hierarchical beam search methods~\cite{li2019explore,wei2019two,khordad2021multiuser}. These approaches reduce scanning overhead by first performing a coarse wide-sector search and then refining the search to identify a narrow transmit--receive beam pair. However, the considered approaches typically assume prior time-frequency synchronization between the \ac{BS} and \ac{UE} or the availability of an out-of-band feedback channel. Such prior coordination cannot be assumed during blind initial acquisition in standalone MIMO-OFDM systems. Although recent compressive sensing frameworks jointly perform synchronization and channel estimation at low \ac{SNR}, their computational requirements hinder practical real-time hardware implementation~\cite{rodriguez2020broadband}. Similarly, AI-based beam training often relies on prior training or beam-quality feedback~\cite{saghezchi-ai-native-ran}, which are unavailable in standalone systems.

Another line of research explores beam acquisition without relying on a feedback channel. In ad hoc networks, for instance, some studies propose blind and asynchronous directional discovery protocols to establish initial beam alignment \cite{wang2021hdnd, hang2024msand}. While these methods eliminate the need for prior synchronization and feedback, they typically rely on randomized beam scanning patterns. This randomness leads to non-deterministic discovery times, making deterministic latency guarantees and practical implementation more challenging. To reduce exhaustive search complexity, recent protocols such as SmartIA~\cite{yazdnian2026smartia} achieve feedback-free beam acquisition where the latency scales logarithmically with the codebook size. However, this relies on multi-peak radiation patterns. Generating these multi-peak patterns requires specialized dual-polarized arrays and complex hybrid precoding. If attempted on a single-RF-chain phased array, synthesizing multiple main lobes inherently splits the transmit power across a wider angular spread, severely degrading the directional gain and received \ac{SNR}.

From an implementation perspective, several \ac{SDR}-based testbeds have recently demonstrated mmWave beam management capabilities~\cite{danneberg2020usrp, marandi2022experimental, ganji2020beamsurfer}. While these experimental platforms and protocols (such as BeamSurfer~\cite{ganji2020beamsurfer}) provide valuable insights into active link maintenance and beam tracking, they operate under the assumption that the initial beam acquisition has already been completed. Additionally, when beam selection is evaluated, these implementations do not fully account for strict hardware-related overhead, such as physical \ac{TDD} and beam switching latencies during the \ac{IA} and re-acquisition phases. Therefore, standalone MIMO-OFDM systems require a practical protocol that establishes blind \ac{IA} while explicitly mitigating these strict hardware constraints. Furthermore, comparatively little attention has been paid to achieving efficient, low-overhead beam re-acquisition following sudden link failures in practical \ac{SDR}-based implementations, especially considering that standard 5G NR beam failure recovery can take up to 1.28\,s~\cite{ganji2020beamsurfer, 3gpp38213}. Addressing these limitations, our framework uniquely achieves deterministic, blind, and asynchronous beam acquisition by explicitly integrating physical \ac{TDD} and beam-switching latencies into a unified \ac{FSM}, bridging initial acquisition with clock-assisted deterministic re-acquisition through standalone \ac{SDR} validation.

\begin{table}[t!]
    \centering
    \caption{Comparison of beam acquisition protocols.}
    \label{tab:sota_comparison}
    \setlength{\tabcolsep}{3pt} 
    \begin{tabular}{l c c c c c}
        \toprule
        \textbf{Work} & \textbf{Standalone$^1$} & \textbf{Determ.$^2$} & \textbf{HW Lat.$^3$} & \textbf{Re-Acq.$^4$} & \textbf{SDR$^5$} \\
        \midrule
        Wang \cite{wang2021hdnd}           & \checkmark & $\times$   & $\times$   & $\times$   & $\times$ \\
        Hang \cite{hang2024msand}          & \checkmark & $\times$   & $\times$   & $\times$   & $\times$ \\
        SmartIA \cite{yazdnian2026smartia} & \checkmark & \checkmark & $\times$   & $\times$   & $\times$ \\
        B.Surfer \cite{ganji2020beamsurfer}& $\times$   & \checkmark & $\times$   & \checkmark & \checkmark \\
        5G NR \cite{3gpp38213}             & $\times$   & \checkmark & \checkmark & \checkmark & $\times$ \\
        \textbf{Proposed}                      & \textbf{\checkmark} & \textbf{\checkmark} & \textbf{\checkmark} & \textbf{\checkmark} & \textbf{\checkmark} \\
        \bottomrule
        \multicolumn{6}{p{0.95\columnwidth}}{\scriptsize $^1$ Fully blind, asynchronous, and feedback-free.} \\
        \multicolumn{6}{p{0.95\columnwidth}}{\scriptsize $^2$ Deterministic beam acquisition time.} \\
        \multicolumn{6}{p{0.95\columnwidth}}{\scriptsize $^3$ Considers \ac{TDD} and beam switching latencies.} \\
        \multicolumn{6}{p{0.95\columnwidth}}{\scriptsize $^4$ Supports rapid clock-assisted re-acquisition after link drop.} \\
        \multicolumn{6}{p{0.95\columnwidth}}{\scriptsize $^5$ Experimental validation on SDR and hardware testbeds.}
    \end{tabular}
\end{table}

\section{System Model}
\label{sec:System_Model}
\begin{figure*}[t!]
    \centering
    \includegraphics[width=\textwidth]{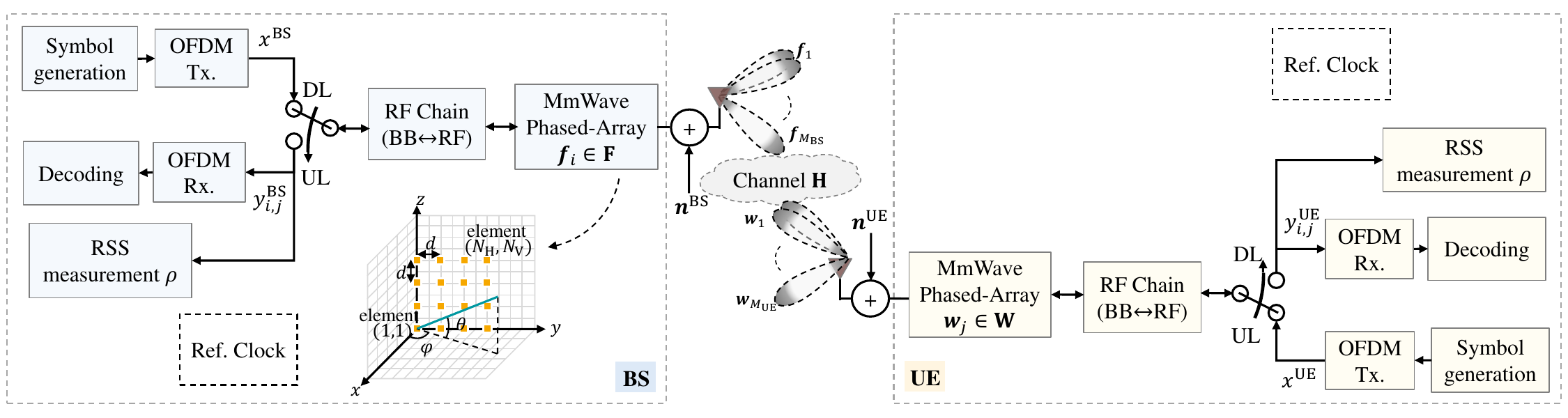}
    \caption{Our considered \ac{mmWave} transceiver architecture, including the TDD-based signal flow and the analog beamformers ($\mathbf{F}, \mathbf{W}$). Note: Additive noise vectors are only applicable during the RX phases.}
    \label{fig:system_model}
\end{figure*}

The physical layer of the proposed standalone MIMO-OFDM system relies on half-duplex transceiver architectures at both the \ac{BS} and \ac{UE}. To support analog beamforming, each node features an electronically steerable \ac{UPA} driven by a single \ac{RF} chain. As illustrated by the transceiver block diagram in Fig.~\ref{fig:system_model}, baseband signals and \ac{RSS} measurements are routed through a \ac{TDD} switch, enabling the nodes to alternate between transmission and reception modes. Consequently, this hardware configuration restricts the system to sequential, time-slotted beam sweeping, which forms the foundation of the protocols detailed later in Section~\ref{sec:Proposed_Algo}.

\subsection{Beamformer}
Antennas on both sides include $N_\mathrm{H}$ elements on the $y$-axis and $N_\mathrm{V}$ elements on the $z$-axis, yielding $N = N_\mathrm{H} N_\mathrm{V}$ elements arranged in a planar grid, as illustrated in Fig.~\ref{fig:system_model}, with uniform inter-element spacing $d$ in both horizontal and vertical directions. Beam steering occurs electronically through applying the complex weights to individual array elements. Such weights are computed using the array factors for the \ac{BS} and \ac{UE} antennas. This array factor is multiplied by the radiation pattern of a single element to yield the complete directive response of the array. For an incident plane wave arriving from elevation angle $\theta$ and azimuth $\phi$, we represent the array factor as $a(\phi,\theta)$.

Inspired by the model in \cite{aykin2020efficient} and the coordinate system from \cite{Bjornson2024}, the signal at element $(h, v)$, denoted $s_{h,v}$, is expressed as
\begin{equation}
    s_{h,v} = g e^{\frac{\mathrm{j}2 \pi \Delta r_{h,v}}{\lambda}},
    \label{eq:rx_sig_smn}
\end{equation}
with $g$ denoting the constant element gain, $\lambda$ denoting the wavelength of the signal, and $\Delta r_{h,v} = ((h-1) \sin \phi \cos \theta + (v-1) \sin \theta)d$ with $h \in \{1, ..., N_\mathrm{H}\}$ and $v \in \{1, ..., N_\mathrm{V}\}$. The output of the antenna array becomes
\begin{equation}
    s =\sum_{h=1}^{N_H} \sum_{v=1}^{N_V} s_{h,v} b_{h,v} = g\sum_{h=1}^{N_H} \sum_{v=1}^{N_V} b_{h,v} e^{\frac{\mathrm{j}2 \pi \Delta r_{h,v}}{\lambda}} = g \;a(\phi,\theta),
    \label{eq:ant_output}
\end{equation}
with $b_{h,v}$ as the beamforming weight for that element (representing the electronic phase shift applied independently of the passive element gain $g$). Under uniform signal amplitudes across antennas, power at the \ac{UPA} maximizes in a given direction by optimizing $|a(\phi,\theta)|$, achieved by choosing $b_{h,v}$ to align phases of all received signals, specifically $b_{h,v} = e^{-\frac{\mathrm{j}2 \pi \Delta r_{h,v}}{\lambda}}$. Thus, the beam directs toward $(\phi, \theta)$, as visualized by the array geometry inset in Fig.~\ref{fig:system_model}.

The sparse \ac{mmWave} link between the \ac{BS} and the \ac{UE} is captured by the downlink channel matrix $\mathbf{H} \in \mathbb{C}^{N \times N}$ (with $\mathbf{H}^T$ denoting the reciprocal uplink channel). To express the received signal, the scalar element weights $b_{h,v}$ are vectorized to form the steering vectors for the antenna arrays. These vectors are precomputed for various angles and stored in a codebook. Let $\mathbf{F} = \{\mathbf{f}_1, \mathbf{f}_2, \ldots, \mathbf{f}_{M_{\mathrm{BS}}}\}$ denote the predefined codebook of the \ac{BS} beamformer, with $\mathbf{f}_i \in \mathbb{C}^{N \times 1}$. Similarly, let $\mathbf{W} = \{\mathbf{w}_1, \mathbf{w}_2, \ldots, \mathbf{w}_{M_{\mathrm{UE}}}\}$ denote the predefined codebook of the \ac{UE} beamformer, with $\mathbf{w}_j \in \mathbb{C}^{N \times 1}$. Here, $M_{\mathrm{BS}}$ and $M_{\mathrm{UE}}$ indicate the total number of predefined beams at each respective node. We assume normalized vectors $\lVert \mathbf{f}_i\rVert_2^2=\lVert \mathbf{w}_j\rVert_2^2=1$.

In reception mode, the complex baseband signals at the \ac{BS} and \ac{UE} for a given beam pair $(i, j)$ are expressed as
\begin{subequations} \label{eq:rx_signal}
\begin{align}
    y_{i,j}^{\mathrm{BS}}(t) &= \mathbf{f}_i^H \mathbf{H}^T \mathbf{w}_j x^{\mathrm{UE}}(t) + \mathbf{f}_i^H \mathbf{n}^{\mathrm{BS}}(t), \label{eq:rx_signal_bs} \\
    y_{i,j}^{\mathrm{UE}}(t) &= \mathbf{w}_j^H \mathbf{H} \mathbf{f}_i x^{\mathrm{BS}}(t) + \mathbf{w}_j^H \mathbf{n}^{\mathrm{UE}}(t), \label{eq:rx_signal_ue}
\end{align}
\end{subequations}
where $x^{\mathrm{BS}}(t)$ and $x^{\mathrm{UE}}(t)$ are the transmitted signals by the \ac{BS} and \ac{UE}, respectively, and $\mathbf{n}^{\mathrm{BS}}(t), \mathbf{n}^{\mathrm{UE}}(t)\sim \mathcal{CN}(\mathbf{0}, \sigma^2\mathbf{I}_N)$ represent the additive white Gaussian noise. During the beam acquisition process, the transceivers measure the power of these received signals over a specified duration to compute the \ac{RSS}, which is then evaluated against the detection threshold. Protocol details are discussed in Section~\ref{sec:Proposed_Algo}.

\subsection{Received Signal Strength Measurement}
Let $t_{d}$ denote the dwell time, i.e., the duration a single beam is held fixed for a reliable measurement. Let \(\tau \in [0,\,t_{d}]\) denote the local time variable within this dwell period. To ensure the measured power corresponds only to the currently active beam and is not corrupted by hardware switching latencies, we compute the \ac{RSS}, \(\rho\), as the time-averaged received energy over the dwell time, explicitly excluding a hardware guard time \(t_{g}\):
\begin{equation}
    \rho = \frac{1}{t_{d} - t_g} \int_{t_g}^{t_{d}} \big|y(\tau)\big|^{2} \, \mathrm{d}\tau,
    \label{eq:rho_calc}
\end{equation}
where $y(\tau)$ corresponds to the received complex baseband signal (defined in \eqref{eq:rx_signal}), expressed in the local time reference of the current beam dwell time. Here, $t_{d}$ represents the beam dwell time for the \ac{UE} in \ac{DL} mode or the \ac{BS} in \ac{UL} mode. The guard interval must satisfy $t_{g}>\max\!\big(t_{\mathrm{sw}}^{\mathrm{BS}}, t_{\mathrm{sw}}^{\mathrm{UE}}\big)$, where $t_{\mathrm{sw}}^{\mathrm{BS}}$, and $t_{\mathrm{sw}}^{\mathrm{UE}}$ are the beam switching latencies for the \ac{BS} and \ac{UE}, respectively.

\section{Standalone mmWave Beam Management Framework}
\label{sec:Proposed_Algo}
\begin{figure}[!t]
    \centering
    \begin{tikzpicture}[
        shorten >=1pt, 
        node distance=2.2cm, 
        on grid, 
        auto, 
        thick,
        scale=0.9, 
        transform shape, 
        every state/.style={
            fill=blue!5, 
            text width=1.5cm, 
            align=center, 
            font=\footnotesize, 
            inner sep=1pt
        },
        every node/.style={font=\scriptsize}
    ]
        
        \node[state, initial, initial text=] (S0) {State 0:\\ Disconnected};
        \node[state] (S1) [right=4.2cm of S0] {State 1:\\ Linked\\ (OFDM Data TX)};
        \node[state] (S2) [below=2.2cm of S1] {State 2:\\ Re-Acquisition};
    
        \path[->] 
            (S0) edge node [align=center] {BIBA $\rightarrow$ Sync\\ Sequence} (S1)
            (S1) edge [bend left=40] node [right, xshift=4pt] {Link Drop} (S2)
            (S2) edge [bend left=40] node [left, xshift=-4pt] {CARA Success} (S1)
            (S2) edge [bend left=50] node [below left] {Timeout/Failure} (S0);
    \end{tikzpicture}
    \caption{\ac{FSM} diagram of the proposed framework}
    \label{fig:state_machine}
\end{figure}

\begin{figure*}
    \centering
    \includegraphics[width=\textwidth]{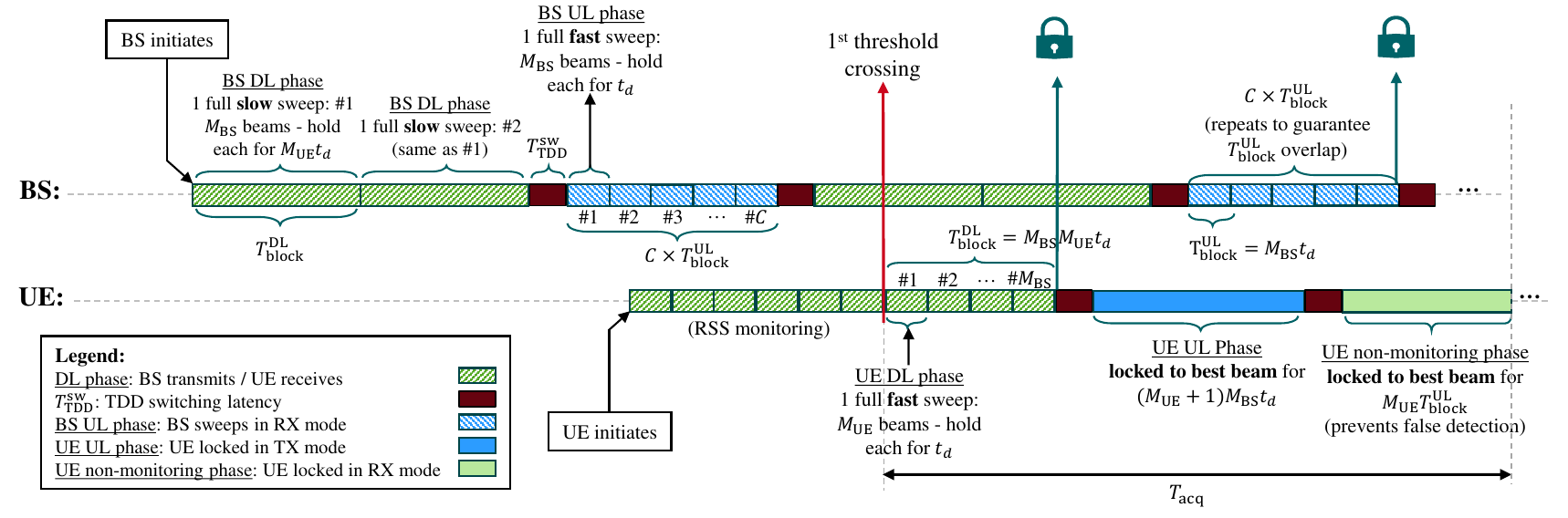}
    \caption{Timing diagram of the \ac{BIBA} protocol~\cite{abdi2026vtc}, illustrating threshold-triggered, blind, and asynchronous \ac{IA}. Due to the uncoordinated start times of the two nodes, the diagram shows one possible relative timing scenario, in which the \ac{BS} initiates the protocol before the \ac{UE}. The fixed-length variation defined in \eqref{eq:c_var_fix} is illustrated with $M_{\mathrm{BS}}=M_{\mathrm{UE}}=4$ and $C=5$.}
    \label{fig:fig_biba}
\end{figure*}

This section details the proposed standalone beam management framework, starting with its unified finite state machine and followed by the specific operations of the \ac{BIBA} and \ac{CARA} protocols.

\subsection{Unified Finite State Machine (FSM) Architecture}
The proposed framework manages the operational states (\ac{IA} and subsequent beam re-acquisition) of directional wireless links through a structured \ac{FSM}, as shown in Fig.~\ref{fig:state_machine}. Initially, the system is in a disconnected state (State 0). First, the \ac{BS} and \ac{UE}, asynchronously, perform \ac{IA} to identify the optimal beam pair using the \ac{BIBA} protocol. Subsequently, with the high \ac{SNR} attained in the \ac{IA} process, a reliable time and frequency synchronization is achieved \cite{Schmidl1997}, enabling OFDM-based data transmission (State 1). To maintain hardware robustness, the link is continuously monitored; if the \ac{RSS} falls below a threshold, the system enters a re-acquisition state (State 2). Here, the \ac{CARA} protocol is triggered. While \ac{BIBA} has an exhaustive search nature due to the lack of an out-of-band feedback channel and prior synchronization, \ac{CARA} exploits the established time synchronization during \ac{IA} to rapidly recover the link, significantly reducing latency compared to a fully blind and asynchronous \ac{IA}. If the link cannot be recovered before the accumulated clock drift exceeds the hardware tolerance, a timeout triggers a hard reset to State 0.

\subsection{Blind and Asynchronous Initial Beam Acquisition (BIBA)}
The proposed \ac{BIBA} protocol, introduced in our preliminary study~\cite{abdi2026vtc},  consists of two distinct probing phases: i) a \ac{DL} phase, in which the \ac{BS} operates in \ac{TX} mode and transmits probing signals, while the \ac{UE} operates in \ac{RX} mode to identify its optimal receive beam; and ii) an \ac{UL} phase, in which the \ac{UE} operates in \ac{TX} mode and transmits probing signals, while the \ac{BS} operates in \ac{RX} mode to identify its optimal receive beam. Since the initial start times of the \ac{BS} and \ac{UE} are completely uncoordinated, due to their asynchronous operation and the absence of an out-of-band feedback channel, different relative time offsets may occur between them when the \ac{UE} starts monitoring the \ac{RSS}. Fig.~\ref{fig:fig_biba} illustrates one such time-offset realization as an example, showing how the \ac{UE} and \ac{BS} identify and lock onto their best beam pair under this specific timing offset.

\subsubsection{DL Phase}
The start time of the protocol at two nodes is independent, i.e., each node is uncertain whether the other node has already started the process or not. The protocol always starts with the \ac{DL} phase at both ends. The \ac{UE} begins its \ac{DL} phase in \ac{RX} mode by sweeping through all possible beam directions $\phi_j$, $j\in[1:M_{\mathrm{UE}}]$ with a $t_{d}$ dwell time (which accounts for both the beam switch latency $t_{\mathrm{sw}}^{\mathrm{UE}}$ and the signal measurement time), while measuring the average \ac{RSS} values for each angle continuously, using \eqref{eq:rho_calc}, and comparing it to the detection threshold value $\gamma$. The time to complete one full \ac{UE} sweep is $M_{\mathrm{UE}}t_{d}$. \ac{UE} repeats sweeping until the first \ac{RSS} threshold crossing happens, confirming the initiation of the \ac{BS} probing phase. However, to guarantee finding the global maximum rather than just the first signal above the threshold, the \ac{UE} cannot yet decide on the best beam until it continues sweeping and measuring the \ac{RSS} ($\rho$) for another $M_{\mathrm{BS}}$ full \ac{UE} sweeps ($T_{\mathrm{block}}^{\mathrm{DL}}=M_{\mathrm{BS}}M_{\mathrm{UE}}t_{d}$, the time between the vertical red arrow and the first lock symbol in Fig.~\ref{fig:fig_biba}), ensuring the \ac{UE} has enough time to probe every possible \ac{BS} transmit beam. This is necessary because the \ac{BS} holds each of its transmit beams for a duration of $M_{\mathrm{UE}}t_{d}$, meaning the \ac{UE} must sweep its entire codebook $M_{\mathrm{BS}}$ times to probe all possible beam pairs between the \ac{BS} and \ac{UE}. After that, the optimal beam index for the \ac{UE} associated with beam direction $\phi_{\mathrm{UE}}^\star=\phi_{\mathrm{j^\star}}$ is found using,
\begin{equation}
j^\star = \argmax_{j\in[1:M_{\mathrm{UE}}]} \left( \max_{i\in[1:M_{\mathrm{BS}}]} \rho_{i,j} \right),
\label{eq:j_star}
\end{equation}
where $\rho_{i,j}$ is the \ac{RSS} value measured when the \ac{BS} transmits with its $i$-th beam and the \ac{UE} receives with its $j$-th beam, computed via \eqref{eq:rho_calc}. This milestone is denoted by the first lock symbol in Fig.~\ref{fig:fig_biba}, indicating the \ac{UE} has successfully locked to its best beam $\phi_{\mathrm{UE}}^\star$. Immediately after this, the \ac{UE} switches to the \ac{UL} phase.

The \ac{BS} independently and asynchronously initiates its \ac{DL} phase in \ac{TX} mode, sweeping its entire beam codebook $\phi_i$, $i\in[1:M_{\mathrm{BS}}]$, twice to ensure that, despite the uncoordinated start times, the \ac{UE} can observe at least one complete \ac{BS} codebook sweep after detecting the first \ac{RSS} threshold crossing. The \ac{BS} holds each transmit beam for a duration of $M_{\mathrm{UE}}t_d$, allowing the \ac{UE} to complete a full sweep of its receive codebook within each \ac{BS} beam dwell interval. Consequently, the total duration of the \ac{BS} probing phase is $2T_{\mathrm{block}}^{\mathrm{DL}}=2M_{\mathrm{BS}}M_{\mathrm{UE}}t_d$. Assuming successful threshold detection, this duration provides the \ac{UE} with sufficient time to probe all possible transmit--receive beam pairs and lock onto its best observed receive beam. The \ac{BS} then transitions to the \ac{UL} phase without requiring feedback or further synchronization.

\begin{algorithm}[t]
\caption{BIBA: BS-side}
\label{alg:bs_acq}
\begin{algorithmic}[1]
\Require $M_{\mathrm{BS}},\;M_{\mathrm{UE}},\;\phi_i,\;t_{d},\;T_{\mathrm{TDD}}^{\mathrm{sw}},\;\gamma,\;C$
\Ensure $i^\star$
\State Enter DL mode and wait $T_{\mathrm{TDD}}^{\mathrm{sw}}$ \Comment{\ac{TX} mode}
\State Initialize: $i^\star \gets \texttt{null}$; \, $\rho_{\mathrm{max}} \gets -\infty$
\While{$i^\star = \texttt{null}$}
    \State $t \gets 0$ \Comment{reset timer}
    \State $\rho_{\mathrm{max}} \gets -\infty$ \Comment{reset max. RSS value}
    \For{$b = 1$ to $2M_{\mathrm{BS}}$}
        \State $i \gets \big((b-1) \bmod M_{\mathrm{BS}}\big) + 1$; Set to $i$th beam ($\phi_i$)
        \State Wait for $M_{\mathrm{UE}}t_{d}$
    \EndFor
    \State Enter UL mode and wait $T_{\mathrm{TDD}}^{\mathrm{sw}}$ \Comment{\ac{RX} mode}
    \State $t \gets t+2M_{\mathrm{BS}}M_{\mathrm{UE}}t_{d}+T_{\mathrm{TDD}}^{\mathrm{sw}}$
    \For{$b = 1$ to $C M_{\mathrm{BS}}$}
        \State $i \gets \big((b-1) \bmod M_{\mathrm{BS}}\big) + 1$; Set to $i$th beam ($\phi_i$)
        \State $\rho \gets \textproc{compute RSS using \eqref{eq:rho_calc}}$
        \If{$\rho \geq \gamma$ \textbf{and} $\rho \geq \rho_{\mathrm{max}}$}
            \State $\rho_{\mathrm{max}} \gets \rho$; \, $i^\star \gets i$
        \EndIf
        \State $t \gets t + t_{d}$
    \EndFor
    \State Enter DL mode and wait $T_{\mathrm{TDD}}^{\mathrm{sw}}$ \Comment{\ac{TX} mode}
    \State $t \gets t+C M_{\mathrm{BS}} t_{d}+T_{\mathrm{TDD}}^{\mathrm{sw}}$
\EndWhile
\State Set the beam towards $\phi_{\mathrm{BS}}^\star=\phi_{i^\star}$
\State \Return $i^\star$
\end{algorithmic}
\end{algorithm}

\begin{algorithm}[t]
\caption{BIBA: UE-side}
\label{alg:ue_acq}
\begin{algorithmic}[1]
\Require $M_{\mathrm{BS}},\;M_{\mathrm{UE}},\;\phi_j,\;t_{d},\;T_{\mathrm{TDD}}^{\mathrm{sw}},\;\gamma$
\Ensure $j^\star$
\State Enter DL mode and wait $T_{\mathrm{TDD}}^{\mathrm{sw}}$ \Comment{\ac{RX} mode}
\State Initialize: $j^\star\gets\texttt{null}$; \, $f_\gamma\gets\texttt{False}$ (threshold crossing flag); \, $b\gets 1$; \, $p\gets 0$; \, $r_{\max}\gets -\infty$; \, $t \gets 0$
\While{$\texttt{True}$}
    \State $j \gets \big((b-1) \bmod M_{\mathrm{UE}}\big) + 1$; Set to $j$th beam ($\phi_j$)
    \State $\rho \gets \textproc{compute RSS using \eqref{eq:rho_calc}}$
    \If{$\rho \ge \gamma$}
        \If{$f_\gamma = \texttt{False}$}
            \State $f_\gamma \gets \texttt{True}$ \Comment{set flag for the 1st detection}
        \EndIf
        \If{$f_\gamma$ \textbf{and} $\rho \geq \rho_{\mathrm{max}}$ \textbf{and} $p < M_{\mathrm{BS}}M_{\mathrm{UE}}$}
            \State $\rho_{\mathrm{max}} \gets \rho$; \, $j^\star \gets j$
        \EndIf
    \EndIf
    \If{$f_\gamma$}
        \State $p\gets p+1$ \Comment{increment post-detection counter}
        \If{$p = M_{\mathrm{BS}}M_{\mathrm{UE}}$} \State \textbf{break} \EndIf
    \EndIf
    \State $b \gets b + 1$ \Comment{increment counter}
    \State $t \gets t +  t_{d}$
\EndWhile
\State $t \gets t + M_{\mathrm{BS}}M_{\mathrm{UE}} t_{d}$
\State Enter UL mode and wait $T_{\mathrm{TDD}}^{\mathrm{sw}}$ \Comment{\ac{TX} mode}
\State Set the beam towards $\phi_{\mathrm{UE}}^\star=\phi_{j^\star}$
\State Wait for $M_{\mathrm{BS}}(M_{\mathrm{UE}}+1)t_{d}$
\State Enter DL mode and wait $T_{\mathrm{TDD}}^{\mathrm{sw}}$ \Comment{\ac{RX} mode}
\State Wait for $M_{\mathrm{BS}}M_{\mathrm{UE}}t_{d}$ \Comment{non-monitoring phase interval}
\State \Return $j^\star$
\end{algorithmic}
\end{algorithm}

\subsubsection{UL Phase}
After a guard interval $T_{\mathrm{TDD}}^{\mathrm{sw}}$, which accounts for hardware \ac{TDD} switching latencies, the \ac{UE} initiates its \ac{UL} phase in \ac{TX} mode, transmitting continuously using its previously selected beam $\phi_{\mathrm{UE}}^\star$ for a duration of $(M_{\mathrm{UE}}+1)M_{\mathrm{BS}}t_d$. As illustrated in Fig.~\ref{fig:fig_biba}, this duration corresponds to $M_{\mathrm{UE}}+1$ consecutive transmission blocks (shown in blue), each of length $T_{\mathrm{block}}^{\mathrm{UL}}=M_{\mathrm{BS}}t_d$. This extended transmission duration accommodates the asynchronous operation of the two nodes, providing sufficient overlap with the \ac{BS}'s receive beam sweeps despite their uncoordinated start times. In the absence of a feedback channel, the \ac{UE} assumes that the \ac{BS} has completed beam selection by the end of this interval (if this assumption fails, e.g., due to sudden blockage or high mobility, the subsequent lack of a detectable signal simply triggers an \ac{FSM} timeout, resetting the protocol to State~0). The \ac{UE} then returns to \ac{DL} mode to eventually monitor for link drops. However, because the \ac{BS} might still be finishing its receive beam sweeps, the \ac{UE} temporarily suspends its \ac{RSS} monitoring to prevent false link drop detection. As indicated in Fig.~\ref{fig:fig_biba}, this non-monitoring phase lasts for an interval of $M_{UE}T_{\mathrm{block}}^{\mathrm{UL}}$, guaranteeing the \ac{BS} has returned to transmission mode before the \ac{UE} expects a stable signal.

From the \ac{BS} perspective, after completing its \ac{DL} phase and the guard interval $T_{\mathrm{TDD}}^{\mathrm{sw}}$, the \ac{BS} independently initiates its \ac{UL} phase in \ac{RX} mode. Due to the asynchronous operation, the \ac{UE} may not yet be transmitting when the \ac{BS} begins its \ac{UL} phase. Nevertheless, the protocol timing is designed to ensure sufficient overlap between the \ac{UE}'s transmission and the \ac{BS}'s receive beam sweeps. The \ac{BS} performs an \ac{RSS} measurement procedure analogous to the \ac{UE}-side \ac{DL} phase, sweeping through its entire receive beam codebook with a dwell time $t_d$ per beam. Each full sweep therefore requires $T_{\mathrm{block}}^{\mathrm{UL}}=M_{\mathrm{BS}}t_d$. Based on the measured \ac{RSS} values, the \ac{BS} selects the receive beam $\phi_{\mathrm{BS}}^\star=\phi_{i^\star}$ associated with the maximum observed \ac{RSS}, where
\begin{equation}
i^\star = \argmax_{i\in[1:M_{\mathrm{BS}}]} \rho_i,
\label{eq:i_star}
\end{equation}
and $\rho_i$ denotes the maximum \ac{RSS} observed for the $i$-th \ac{BS} receive beam during the measurement interval. This beam selection is indicated by the second lock symbol in Fig.~\ref{fig:fig_biba}, marking the completion of the \ac{BS}-side beam selection procedure.

To improve acquisition robustness, the \ac{BS} may perform multiple ($C$) complete receive beam sweeps during the \ac{UL} phase. Increasing $C$ extends the observation interval, reducing the risk of insufficient \ac{TX}--\ac{RX} overlap at the cost of a proportionally longer acquisition time. This introduces a trade-off between acquisition reliability and latency. In this work, two design variations are considered:
\begin{equation}
C =
\begin{cases}
1, & \text{variable length},\\[3pt]
M_{\mathrm{UE}}+1, & \text{fixed length}.
\end{cases}
\label{eq:c_var_fix}
\end{equation}
Here, $C$ denotes the number of complete \ac{BS} receive beam sweeps during the \ac{UL} phase, represented by the blue blocks in Fig.~\ref{fig:fig_biba}, which illustrates the case $C=5$. In the fixed-length variation ($C=M_{\mathrm{UE}}+1$), the \ac{BS} performs multiple sweeps to accommodate the asynchronous start times, ensuring sufficient \ac{TX}--\ac{RX} overlap for beam selection under the assumed detection and timing conditions. This provides a fixed acquisition duration at the cost of increased latency. In contrast, the variable-length variation ($C=1$) limits the \ac{BS} to a single sweep per \ac{UL} phase, reducing the duration of each acquisition attempt but introducing a non-zero probability of acquisition failure due to insufficient \ac{TX}--\ac{RX} overlap. Consequently, the overall acquisition time varies depending on the relative start times of the two nodes and the number of acquisition attempts required.

The \ac{BIBA} protocol operates independently of which node initiates the process first. Protocols~\ref{alg:bs_acq} and \ref{alg:ue_acq} describe the \ac{BS} and \ac{UE} operations as two independent processes to facilitate practical implementation. Until the initial beam acquisition procedure is successfully completed, the \ac{BS} repeatedly executes its \ac{DL}--\ac{UL} routine, while the \ac{UE} continuously sweeps its receive beams in the \ac{DL} phase until the first \ac{RSS} threshold crossing occurs. Upon successful completion of \ac{BIBA}, the nodes establish a communication link for data transmission and achieve wall-clock synchronization, which is subsequently leveraged for fast beam re-acquisition.

\begin{figure}
    \centering
    \includegraphics[width=0.42\textwidth]{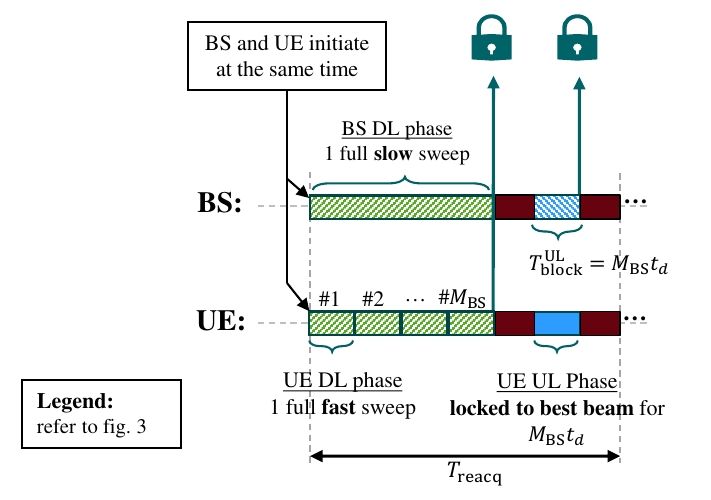}
    \caption{Timing diagram for the \ac{CARA} protocol, illustrating synchronized re-acquisition following an RSS drop. Unlike \ac{BIBA}, in \ac{CARA}, the nodes leverage their previously established wall-clock synchronization to align their \ac{DL} and \ac{UL} sweep phases, resolving the initial asynchronous timing ambiguity. This drastically reduces the overall re-acquisition time.}
    \label{fig:fig_cara}
\end{figure}

\subsection{Clock-Assisted Re-Acquisition (CARA)}
Once \ac{IA} is completed and the time synchronization of the \ac{BS} and \ac{UE} is finished, data transmission begins. During this phase, the receiver continuously monitors the link quality, and a re-acquisition is triggered if the \ac{RSS} falls below the threshold $\gamma$, which may occur due to mobility or emergence of a blockage. In such cases, the previously selected beams are no longer optimal.

Unlike the initial \ac{BIBA} phase, the system now leverages the wall-clock synchronization established during the previous data exchange. Because the transceivers are driven by high-stability oscillators (e.g., \ac{SDR} platforms with a specified crystal oscillator stability of $\pm 2.5$~ppm \cite{usrpx410_datasheet}), the accumulated clock drift is bounded to $2.5~\mu\text{s/s}$. Consequently, even after several seconds of total link failure, the timing misalignment remains negligible compared to the hardware guard interval and re-acquisition times. While prior works have conceptually observed that local clock synchronization persists during temporary blockages \cite{ganji2020beamsurfer}, they primarily rely on continuously scanning adjacent beams to maintain the link.

In contrast, the \ac{CARA} protocol formalizes this clock-retention property into a strictly scheduled re-acquisition process. Because both nodes establish a schedule of predefined wall-clock times upon entering State 1, they independently wait for the next scheduled time after a link drop to execute the deterministic \ac{CARA} sweep, as illustrated in Fig.~\ref{fig:fig_cara}. The procedure comprises an exhaustive \ac{DL} sweep over $T_{\mathrm{block}}^{\mathrm{DL}}$, a $T_{\mathrm{TDD}}^{\mathrm{sw}}$ guard interval, and a final \ac{UL} \ac{BS} sweep of $T_{\mathrm{block}}^{\mathrm{UL}}$. By eliminating the redundant, overlapping sweeps required for blind and asynchronous \ac{IA} (\ac{BIBA}), \ac{CARA} significantly reduces the re-acquisition overhead compared to \ac{BIBA}.

\begin{figure*}
    \centering
    \includegraphics[width=\textwidth]{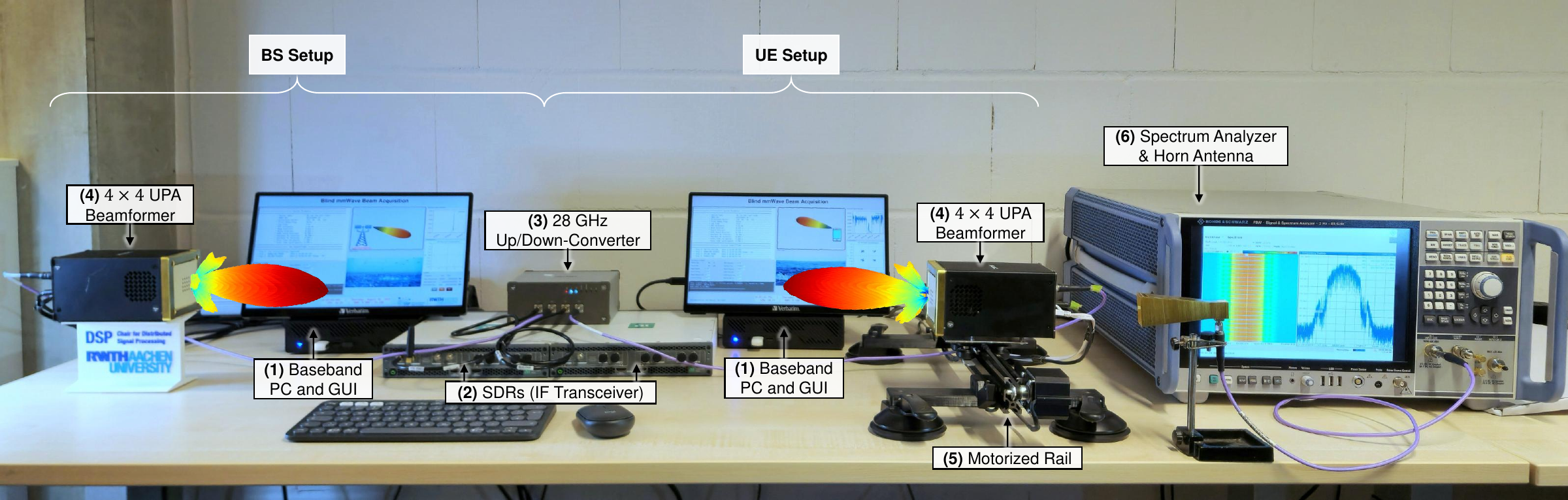}
    \caption{\ac{HW} setup: \ac{BS}/\ac{UE} nodes with \ac{SDR}-based IF transceivers, 28~GHz \ac{RF} front-ends, a motorized rail, and a spectrum analyzer for over-the-air verification. The overlaid 3D projections represent the measured \ac{UPA} radiation patterns.}
    \label{fig:testbed_setup}
\end{figure*}

To guarantee the deterministic \ac{FSM} remains valid, \ac{CARA} enforces a hard mathematical timeout. The protocol relies on the intra-symbol guard time $t_g$ to absorb both \ac{RF} switching transients and accumulated clock drift. If a link drop lasts for a duration $T_{\mathrm{timeout}}$, the maximum accumulated drift at a combined rate of $D_{\mathrm{rate}}$ must not exceed $t_g$. Therefore, the \ac{FSM} forces a transition from State 2 back to State 0 (Disconnected) if the timeout duration exceeds the theoretical boundary:
\begin{equation}
T_{\mathrm{timeout}}^{\mathrm{max}} \le \frac{t_g}{D_{\mathrm{rate}}}.
\label{eq:cara_timeout}
\end{equation}
Given a worst-case combined drift of $D_{\mathrm{rate}} = 5.0~\mu\text{s/s}$ and a testbed intra-symbol guard time of $t_g = 1$~ms, our testbed implementation can sustain a total link loss for a theoretical upper bound of 200 seconds before a full blind and asynchronous \ac{IA} reset is strictly required, though practical protocol timeouts can be configured much shorter.

\section{Experimental Evaluation and Discussion}
\label{sec:Experimental_Evaluation_and_Discussion}
This section evaluates our proposed protocols to validate the contributions. First, Section~\ref{subsec:exp_setup} describes the 28\,GHz testbed used to assess practical hardware constraints. Next, Section~\ref{subsec:software_latency} presents the \ac{SDR} software latency measurements and optimization. Section~\ref{subsec:acq_time} establishes the acquisition times derived from measured hardware latencies for the \ac{BIBA} and \ac{CARA} protocols. Finally, Section~\ref{subsec:mobility} provides a probabilistic analysis to establish the maximum supported \ac{UE} speeds and mobility boundaries.

\subsection{Experimental Setup}
\label{subsec:exp_setup}
The experimental platform is built upon our mmWave MIMO testbed~\cite{lee2025}. The platform is comprised of several integrated components, as shown in Fig.~\ref{fig:testbed_setup}. Two \acp{PC} (1) perform baseband signal processing, real-time beam control, and \ac{GUI} visualization, executing the \ac{BIBA} and \ac{CARA} protocols. Each \ac{PC} interfaces with a dedicated \ac{SDR} device (2) serving as the \ac{IF} transceiver at a 2 GHz frequency. The \ac{IF} signal is then up-converted to the \ac{RF} 28 GHz \ac{mmWave} band by the \ac{mmWave} up/down-converter (3) and routed to two $4\times4$-element \ac{UPA} beamformers (4). These arrays consist of $N = N_\mathrm{H} N_\mathrm{V} = 16$ elements, characterized by a laboratory-measured 3D boresight radiation pattern with a Half-Power Beam-Width (HPBW) of $26.5^{\circ}$, in both azimuth and elevation planes, which is visualized as an overlay on the transceivers in Fig.~\ref{fig:testbed_setup}. The system operates strictly in \ac{TDD} mode, as each node utilizes a single \ac{RF} chain for both transmission and reception. The key system and hardware parameters summarized in Table~\ref{tab:parameters} are extracted directly from the physical component specifications and limitations of the testbed.

A motorized rail (5) controlled by a microcontroller automates the lateral movement of the \ac{UE} to evaluate acquisition under mobility. To independently verify successful beam alignment and monitor post-acquisition link stability, a \ac{mmWave}-capable horn antenna connected to a spectrum analyzer (6) captures the over-the-air signals, visualizing the power spectrum and waterfall of the \ac{OFDM} waveform during live video transmission. The steering directions for the \ac{BS} and \ac{UE} are selected from the following discrete set of azimuth angles:
\begin{equation}
    \phi_{i}, \phi_{j} \in \{-32^\circ, -10^\circ, 10^\circ, 32^\circ\}, \quad i,j \in \{1,2,3,4\}.
    \label{eq:codebook_HWimpl}
\end{equation}
The reduced codebook size ($M_{\mathrm{BS}}=M_{\mathrm{UE}}=4$) uses non-uniform beam spacing to maintain uniform coverage, accommodating hardware and laboratory constraints while enabling experimental validation of the proposed acquisition procedures and their associated hardware latencies. Although practical \ac{mmWave} systems may employ larger codebooks, the acquisition-time expressions derived in the preceding sections characterize the dependence on $M_{\mathrm{BS}}$ and $M_{\mathrm{UE}}$. Furthermore, the \ac{UE} movement is restricted to a linear trajectory facing the \ac{BS}, consistent with the arrays' limited $\pm45^\circ$ electronic steering range. While this configuration does not reproduce arbitrary mobility trajectories, it provides controlled and repeatable conditions for evaluating the protocols under lateral movement. More general mobility conditions are evaluated analytically later.

While the \ac{RSS}-based design inherently allows the transceivers to lock onto the strongest available spatial path, our experimental validation was deliberately conducted in a dominant-\ac{LoS} environment. This controlled setting was chosen to accurately baseline the protocol's fundamental hardware timing performance and \ac{FSM} transition reliability. Accordingly, a static detection threshold $\gamma$ was empirically configured to maintain a reliable margin above the hardware noise floor. Should the protocol lock onto a transient multipath reflection that subsequently decays, the resulting RSS drop inherently triggers an FSM transition to State 2 (CARA) for rapid re-acquisition. A detailed analysis of \ac{SNR}, threshold selection, and multipath effects is left for future work.

\begin{table}[htbp]
\centering
\small
\caption{Key System and Hardware Parameters}
\label{tab:parameters}
\begin{tabular}{lc}
\toprule
\textbf{Parameter} & \textbf{Value} \\
\midrule
Carrier frequency / Array size & 28 GHz / $16 \ (4\times4)$ \\
Codebook size ($M_{\mathrm{BS}}, M_{\mathrm{UE}}$) & 4 \\
Switching dwell time ($t_{d}$) & 3 ms \\
Intra-symbol guard time ($t_g$) & 1 ms \\
TDD switching latency ($T_{\mathrm{TDD}}^{\mathrm{sw}}$) & 10 ms \\
Max clock drift rate ($D_{\mathrm{rate}}$) & 5.0 $\mu$s/s \\
\bottomrule
\end{tabular}
\end{table}

\subsection{\ac{SDR} Software Latency Measurements and Optimization}
\label{subsec:software_latency}
\begin{figure}[t]
    \centering
    \includegraphics[width=0.95\linewidth]{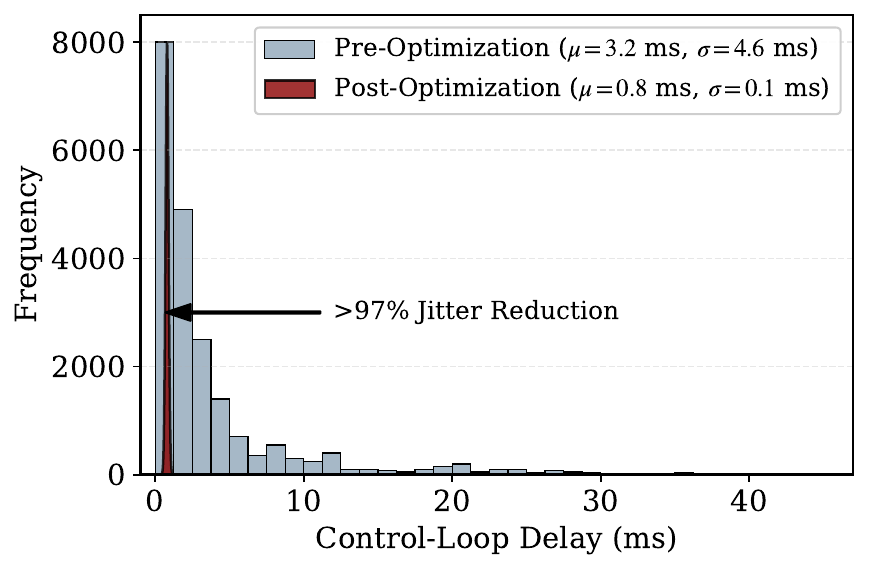}
    \caption{Measured control delay distribution in the \ac{SDR} implementation.}
    \label{fig:software_jitter}
\end{figure}

Executing the protocol \ac{FSM} on a host \ac{PC} introduces non-deterministic \ac{SDR} software latency. As Fig.~\ref{fig:software_jitter} shows, standard \ac{SDR} processing architectures using large buffers and independent software modules cause unpredictable delays, with a severe tail extending beyond 40\,ms. To meet the timing assumptions of both protocols ($t_g > \max(t_{\mathrm{sw}}^{\mathrm{BS}}, t_{\mathrm{sw}}^{\mathrm{UE}})$), we consolidated \ac{RSS} calculation, thresholding, and beam control into a single, minimally buffered block. This optimization reduced the average control-loop delay from 3.2\,ms to 0.8\,ms and the standard deviation by over 97\% (4.6\,ms to 0.1\,ms)~\cite{abdi2026vtc}, bringing the mean delay below the guard time $t_g = 1$\,ms.

\subsection{Beam Acquisition Time}
\label{subsec:acq_time}
Given the input parameter set $\{M_{\mathrm{BS}}, M_{\mathrm{UE}}, t_{d}, T_{\mathrm{TDD}}^{\mathrm{sw}}\}$, the total acquisition times, regardless of which side (\ac{BS} or \ac{UE}) starts first, are as follows:
\begin{itemize}
    \item \textit{\ac{BIBA} protocol}:
    The total beam acquisition time at the \ac{BS}, $T_{\mathrm{acq}}^{\mathrm{BS}}$, is the sum of the initial \ac{DL} phase sweep duration ($2M_{\mathrm{BS}}M_{\mathrm{UE}}t_{d}$), the \ac{UL} phase sweep duration ($C M_{\mathrm{BS}} t_{d}$), and the two required hardware TDD switching latencies ($2T_{\mathrm{TDD}}^{\mathrm{sw}}$). This yields:
    \begin{equation}
        \begin{aligned}
            T_{\mathrm{acq}}^{\mathrm{BS}} &= (2M_{\mathrm{UE}}+C) M_{\mathrm{BS}} t_{d} + 2T_{\mathrm{TDD}}^{\mathrm{sw}} \\
            &= \begin{cases}
                (2M_{\mathrm{UE}}+1) M_{\mathrm{BS}} t_{d} + 2T_{\mathrm{TDD}}^{\mathrm{sw}}, & \text{var. len.}\\[3pt]
                (3M_{\mathrm{UE}}+1) M_{\mathrm{BS}} t_{d} + 2T_{\mathrm{TDD}}^{\mathrm{sw}}, & \text{fix. len.}
               \end{cases}
        \end{aligned}
        \label{eq:T_acq_BS}
    \end{equation}
    where \(C\) is given in \eqref{eq:c_var_fix}. Although the \ac{UE} identifies its beam during the initial \ac{DL} sweep, requiring only $M_{\mathrm{BS}}M_{\mathrm{UE}}t_d$, its total acquisition time also includes the additional fixed slots until the \ac{RSS}-drop monitoring point (see Fig.~\ref{fig:fig_biba}), yielding
    \begin{equation}
        \begin{aligned}
            T_{\mathrm{acq}}^{\mathrm{UE}}&=(3M_{\mathrm{UE}}+1) M_{\mathrm{BS}} t_{d}+2T_{\mathrm{TDD}}^{\mathrm{sw}},\\
        \end{aligned}
        \label{eq:T_acq_UE_2}
    \end{equation}
    that is equal to the acquisition time of the fixed-length version in \eqref{eq:T_acq_BS}. The total acquisition time is defined as the longest duration of the two:
    \begin{equation}
        \begin{aligned}
            T_{\mathrm{acq}}&=\max\!\big(T_{\mathrm{acq}}^{\mathrm{BS}},T_{\mathrm{acq}}^{\mathrm{UE}}\big)\\
            &=(3M_{\mathrm{UE}}+1) M_{\mathrm{BS}} t_{d}+2T_{\mathrm{TDD}}^{\mathrm{sw}}.\\
        \end{aligned}
        \label{eq:T_acq}
    \end{equation}

    \item \textit{\ac{CARA} protocol}:
    It can be seen from Fig.~\ref{fig:fig_cara} that $T_{\mathrm{reacq}}=T_{\mathrm{reacq}}^{\mathrm{BS}}=T_{\mathrm{reacq}}^{\mathrm{UE}}$, where
    \begin{equation}
        \begin{aligned}
            T_{\mathrm{reacq}}&=(M_{\mathrm{UE}}+1) M_{\mathrm{BS}} t_{d}+2T_{\mathrm{TDD}}^{\mathrm{sw}}.\\
        \end{aligned}
        \label{eq:T_reacq}
    \end{equation}
\end{itemize}
The ratio $T_{\mathrm{reacq}}/T_{\mathrm{acq}}$ illustrates the structural efficiency of \ac{CARA}. Assuming the hardware guard interval $T_{\mathrm{TDD}}^{\mathrm{sw}}$ is negligible compared to the sweep durations, this overhead ratio simplifies mathematically to $T_{\mathrm{reacq}}/T_{\mathrm{acq}} \approx (M_{\mathrm{UE}}+1)/(3M_{\mathrm{UE}}+1)$. Applying our constrained hardware parameters ($M_{\mathrm{BS}}=M_{\mathrm{UE}}=4$, $t_{d}=3$\,ms, and $T_{\mathrm{TDD}}^{\mathrm{sw}}=10$\,ms), the total theoretical acquisition times evaluate to 176\,ms for \ac{BIBA} (fixed-length) and 80\,ms for \ac{CARA}. These testbed-limited values would scale down linearly in commercial deployments. Conducting comprehensive statistical timing measurements and experimental validation across diverse mobility conditions are left for future work. By leveraging prior synchronization to eliminate redundant search cycles, \ac{CARA} reduces the theoretical acquisition time by approximately 55\% compared with \ac{BIBA} under the considered hardware configuration. The successful operation of both protocols was also demonstrated on the experimental testbed.

While the 176\,ms acquisition time is practical for the implemented codebook ($M_{\mathrm{BS}}=4$, $M_{\mathrm{UE}}=4$), the exhaustive nature of the \ac{BIBA} protocol results in an $\mathcal{O}(M_{\mathrm{BS}} M_{\mathrm{UE}})$ time complexity. Extending the protocols to larger codebooks (e.g., $M_{\mathrm{BS}}=64, M_{\mathrm{UE}}=16$) would significantly increase latency. To mitigate this scalability bottleneck in commercial systems, future versions of the \ac{FSM} can integrate hierarchical (wide-to-narrow) beam codebooks to establish a rapid coarse connection prior to beam refinement.


\subsection{Mobility Boundaries and Probabilistic Feasibility}
\label{subsec:mobility}
The decision region duration $T_{\phi}$ is defined as the time for the \ac{UE}'s angular position to traverse a single \ac{BS} beam spacing. Equivalently,
\begin{equation}
    T_{\phi} = \frac{\Delta\phi}{\omega_{\mathrm{UE}}} = \frac{\Delta\phi \, D}{v_{\mathrm{UE}}},
\end{equation}
where \(\Delta\phi\) is the boresight-to-boresight angular separation of adjacent beams in radians, $\omega_{\mathrm{UE}}$ and $v_{\mathrm{UE}}$ are the angular and linear speeds of \ac{UE}, respectively, and \(D\) is the distance between the \ac{BS} and the \ac{UE}. Note that this formulation assumes the user moves strictly perpendicular to the \ac{BS}. Because moving directly toward or away from the \ac{BS} causes almost no angular change, this perpendicular assumption represents the most restrictive worst-case scenario for beam acquisition. 

However, in practical deployments, the \ac{UE} trajectory is random. Let $\psi \sim \mathcal{U}[0, 2\pi)$ denote the user's random trajectory angle relative to the \ac{BS} line-of-sight. The angular duration of the decision region is governed exclusively by the perpendicular velocity component, $v_{\perp} = v_{\mathrm{UE}} |\sin\psi|$. Consequently, the decision region becomes a stochastic function of the user trajectory:
\begin{equation}
    T_{\phi}(\psi)=\frac{\Delta\phi \, D}{v_{\mathrm{UE}} |\sin\psi|}.
    \label{eq:decision_region}
\end{equation}

To ensure the system remains efficient, we require the \ac{IA} or re-acquisition process to occupy only a small fraction of the decision region. This is controlled by an efficiency factor $\alpha$:
\begin{equation}
    T_{\phi}(\psi) > \begin{cases}
            \alpha\Big[(3M_{\mathrm{UE}}+1)M_{\mathrm{BS}}\,t_{d}+2T_{\mathrm{TDD}}^{\mathrm{sw}}\Big], & \text{BIBA}\\[3pt]
            \alpha\Big[(M_{\mathrm{UE}}+1)M_{\mathrm{BS}}\,t_{d}+2T_{\mathrm{TDD}}^{\mathrm{sw}}\Big]. & \text{CARA}
       \end{cases}
    \label{eq:feasibility_cond}
\end{equation}

Because this condition depends on the random trajectory $\psi$, we derive the probability of feasible acquisition, $P_{\mathrm{feas}}$, as a closed-form expression (detailed in Appendix~\ref{app:prob_derivation}):
\begin{equation}
    P_{\mathrm{feas}} = \frac{2}{\pi} \arcsin\left( \min\left(1, \frac{\Delta\phi \, D}{\alpha v_{\mathrm{UE}} T_{\mathrm{acq}}} \right) \right).
    \label{eq:prob_acq}
\end{equation}

\begin{figure}[t]
    \vspace{-0.5em}
    \centering
    \includegraphics[width=0.95\linewidth]{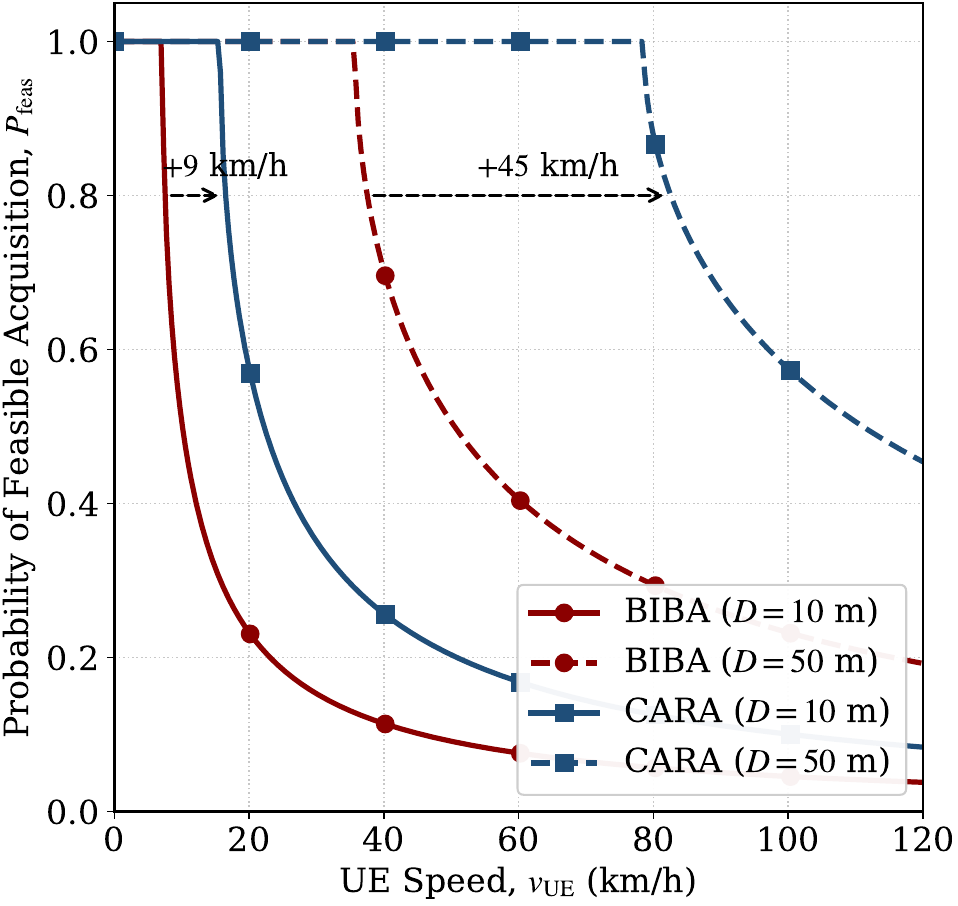}
    \caption{Probability of feasible acquisition ($P_{\mathrm{feas}}$) vs. \ac{UE} speed for random trajectories at $D=10$\,m and $D=50$\,m ($\alpha=10$).}
    \label{fig:prob_acq_curve}
\end{figure}

To visualize, Fig.~\ref{fig:prob_acq_curve} illustrates $P_{\mathrm{feas}}$ for both protocols at fixed distances of 10\,m and 50\,m, and with $\Delta \phi \approx 21^\circ$ for our codebook. At lower speeds, the overhead constraint is satisfied for all possible trajectory angles ($P_{\mathrm{feas}} = 1.0$). As speed increases beyond the critical perpendicular limit, the probability decays, forming sweeping tails where acquisition is only successful if the user is moving at a steep angle toward or away from the \ac{BS}. Because of its significantly lower latency, \ac{CARA} maintains near-certain reliability at much higher speeds, indicating higher resilience to stochastic mobility compared to \ac{BIBA}.

While the worst-case boundary ($P_{\mathrm{feas}}=1.0$) guarantees acquisition feasibility under perpendicular movement, systems may operate under relaxed probability of feasible acquisition targets. By fixing a target probability $P < 1.0$ (assuming the argument of the arcsine function in \eqref{eq:prob_acq} is strictly less than 1), we apply the sine function to isolate the maximum permissible \ac{UE} speed, $v_{\mathrm{UE}}^{\mathrm{max}}$, for any given distance:
\begin{equation}
    v_{\mathrm{UE}}^{\mathrm{max}}(P) = \frac{\Delta\phi \, D}{\alpha T_{\mathrm{acq}} \sin\left(P \frac{\pi}{2}\right)}.
\end{equation}

\begin{figure}[t]
    \vspace{-0.5em}
    \centering
    \includegraphics[width=0.95\linewidth]{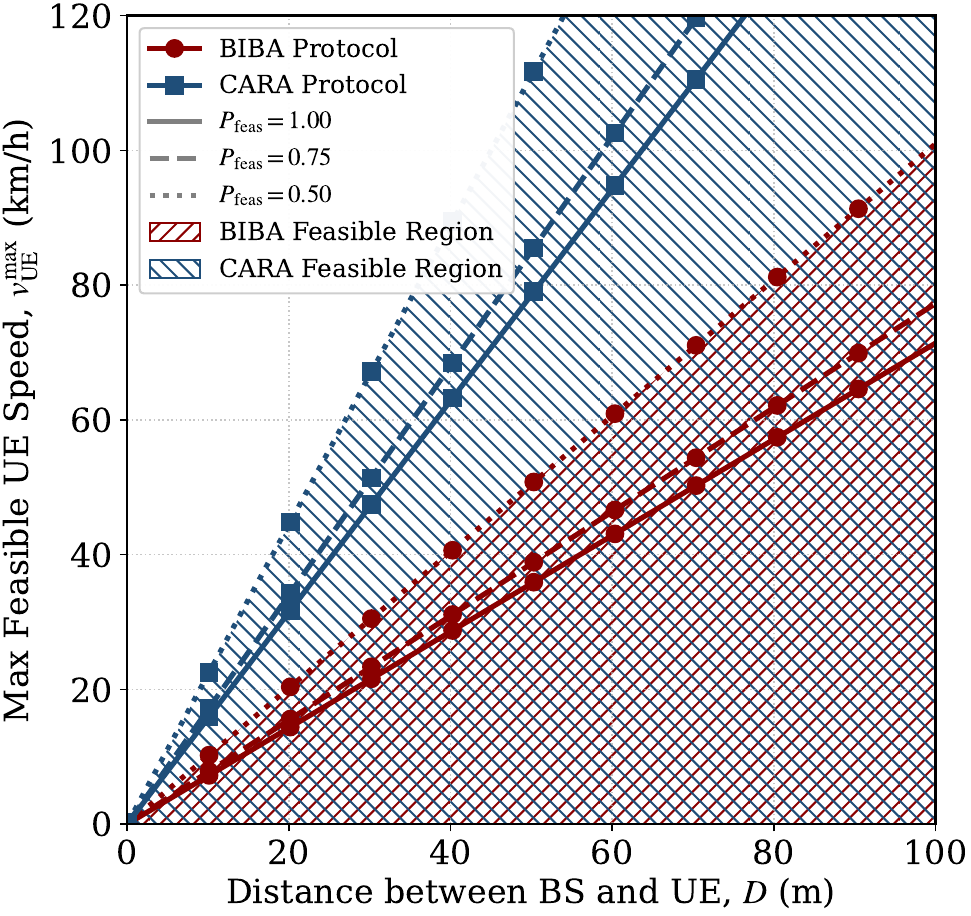}
    \caption{Maximum feasible \ac{UE} speed versus distance for varying target probability of feasible acquisition ($P_{\mathrm{feas}}$) at $\alpha=10$.}
    \label{fig:mobility_boundaries}
\end{figure}

For the feasibility analysis in Fig.~\ref{fig:mobility_boundaries}, we set $\alpha=10$. This choice effectively bounds the beam acquisition overhead to 10\% of the decision interval, ensuring 90\% of the time remains available for subsequent synchronization and \ac{OFDM} data transmission phases as defined in our \ac{FSM} (Fig.~\ref{fig:state_machine}). Based on \eqref{eq:prob_acq} and the velocity bounds, the system exhibits a direct linear sensitivity to $\alpha$: decreasing $\alpha$ (e.g., $\alpha=5$, allowing 20\% overhead) doubles the maximum supported speeds and expands the feasible region, but proportionally reduces the data transmission window. Conversely, increasing $\alpha$ (e.g., $\alpha=20$, restricting overhead to 5\%) guarantees a massive data-transmission interval but halves the mobility tolerance. Therefore, $\alpha=10$ is selected as a representative baseline to balance high-speed vehicular mobility with sufficient data throughput.

Fig.~\ref{fig:mobility_boundaries} plots the maximum supported speeds for both \ac{BIBA} and \ac{CARA} across varying probability thresholds. The solid lines ($P=1.0$) represent the worst-case boundaries. However, if the system can tolerate a lower probability of feasible acquisition target, such as $P=0.75$, the allowable mobility region expands significantly. Because \ac{CARA} inherently requires 55\% less acquisition time, its boundaries are uniformly steeper, demonstrating that it can operate reliably at much higher vehicular speeds compared to \ac{BIBA}.

\section{Conclusions} 
\label{sec:Conclusions}
This work introduced a hardware-robust beam acquisition framework for standalone mobile MIMO-OFDM systems. Validated on a \ac{mmWave} 28\,GHz platform, the integrated \ac{BIBA} and \ac{CARA} protocols deliver reliable, asynchronous, and feedback-free beam acquisition alongside a $\mathord{\sim}55\%$ reduction in re-acquisition overhead. The operational limits of both protocols were established through closed-form probabilistic feasibility bounds under random user mobility trajectories. Furthermore, the underlying \ac{FSM} logic ensures continuous operation by explicitly accommodating hardware-specific impairments such as \ac{TDD} latencies and \ac{SDR} software processing delays. By eliminating the dependency on an external feedback channel and prior synchronization, this framework provides a practical path toward autonomous standalone deployments. A natural extension of this work involves scaling the codebook to massive MIMO arrays (potentially utilizing hierarchical beam structures) and evaluating protocol performance in dense multipath and multi-user environments.

\appendices
\appendices
\section{Derivation of Acquisition Probability}
\label{app:prob_derivation}
Substituting \eqref{eq:decision_region} into the feasibility condition \eqref{eq:feasibility_cond} yields:
\begin{equation}
    \frac{\Delta\phi \, D}{v_{\mathrm{UE}} |\sin\psi|} > \alpha T_{\mathrm{acq}} \implies |\sin\psi| < \frac{\Delta\phi \, D}{\alpha v_{\mathrm{UE}} T_{\mathrm{acq}}}.
\end{equation}
Let $\beta = (\Delta\phi \, D) / (\alpha v_{\mathrm{UE}} T_{\mathrm{acq}})$. The condition simplifies to $|\sin\psi| < \beta$. 

Because $\beta \ge 0$, we consider two cases:
\begin{enumerate}
    \item \textit{Case $\beta \ge 1$}: The condition $|\sin\psi| < \beta$ strictly holds for all trajectory angles. Thus, $P_{\mathrm{feas}} = 1$.
    \item \textit{Case $\beta < 1$}: Due to the symmetry of $|\sin\psi|$, we restrict our analysis to the first quadrant, $\psi \in [0, \pi/2]$. The condition is satisfied for $\psi \in [0, \arcsin(\beta)]$. Because $\psi \sim \mathcal{U}[0, 2\pi)$, the probability is simply the ratio of this valid angular range to the quadrant span:
    \begin{equation}
        P_{\mathrm{feas}} = \frac{\arcsin(\beta)}{\pi/2} = \frac{2}{\pi} \arcsin(\beta).
    \end{equation}
\end{enumerate}
Combining both cases via $\min(1, \beta)$ yields the closed-form expression presented in \eqref{eq:prob_acq}.



\bibliographystyle{IEEEtran}
\bibliography{Bibliography}

\begin{IEEEbiography}[{\includegraphics[width=1in,height=1.25in,clip,keepaspectratio]{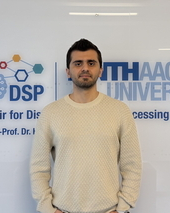}}]{Ali Abdi}
received the Bachelor of Science degree in communications engineering from Shahid Beheshti University, Tehran, Iran, in 2015, and the Master of Science degree in communication systems engineering from Sharif University of Technology, Tehran, Iran, in 2017. He is currently pursuing the Ph.D. degree at RWTH Aachen University in Aachen, Germany. His research interests include channel estimation for 6G, machine learning in wireless PHY, mmWave, and massive MIMO.
\end{IEEEbiography}

\begin{IEEEbiography}[{\includegraphics[width=1in,height=1.25in,clip,keepaspectratio]{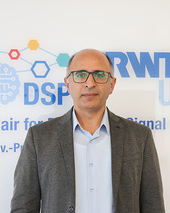}}]{Firooz B. Saghezchi}
(Senior Member, IEEE) received the Ph.D. degree in telecommunications engineering from the University of Aveiro, Portugal, in 2016. From 2010 to 2023, he was a Research Fellow with the Instituto de Telecomunicações, Portugal. Since 2023, he has been with the Chair for Distributed Signal Processing, RWTH Aachen University, Germany, as a Senior Researcher. His research interests include wireless communications, signal processing, machine learning, and radio resource management.
\end{IEEEbiography}

\begin{IEEEbiography}[{\includegraphics[width=1in,height=1.25in,clip,keepaspectratio]{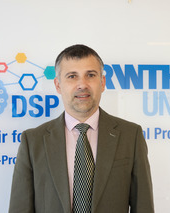}}]{Haris Ga\v{c}anin}
(Fellow, IEEE) received the Ph.D. degree from Tohoku University, Japan, in 2008. He was an Assistant Professor with Tohoku University before joining Alcatel-Lucent Bell (now Nokia Bell Labs) in 2010, where he served as a Physical-Layer Expert and later as Department Head. Since April 2020, he has been a Full Professor at the Chair for Distributed Signal Processing at RWTH Aachen University in Aachen, Germany. His research interests include broad areas of signal processing and machine learning in wireless communications.
\end{IEEEbiography}

\vfill

\end{document}